\documentclass[11pt,twoside]{article}
\usepackage{asp2010}

\resetcounters

\begin{document}

\title{A 3D Radiative Transfer Code for Modeling the Hanle Effect in the Lyman\,$\alpha$ line}
\author{
J.~\v{S}t\v{e}p\'an$^{1,2,3}$, J.~Trujillo~Bueno$^{1,2,4}$
\affil{$^1$Instituto de Astrof\'{\i}sica de Canarias, E-38205 La Laguna, Tenerife, Spain}
\affil{$^2$Departamento de Astrof\'\i sica, Universidad de La Laguna, Tenerife, Spain}
\affil{$^3$Astronomical Institute ASCR, Ond\v{r}ejov, Czech Republic}
\affil{$^4$Consejo Superior de Investigaciones Cient\'\i ficas, Spain}
}

\vspace*{5mm}

In order to obtain empirical information on the magnetism of the solar transition region we need to measure and interpret the linear polarization produced by scattering processes in FUV and EUV spectral lines. Via the Hanle effect such linear polarization signals are sensitive to the magnetic fields expected for the quiet and active regions of the outer solar atmosphere. For example, the Ly$\alpha$ line of H {\sc i} at 1216 \AA\ is mainly sensitive to magnetic strengths between 10 and 100 G, while the Ly$\alpha$ line of He {\sc ii} at 304 \AA\ is mainly sensitive to magnetic strengths between 100 and 1000 G, approximately \citep[see][]{jtblya11a,jtblya11b}. 

The measurement of the Stokes profiles in FUV and EUV lines requires the development of suitable instrumentation, such as the Chromospheric Ly$\alpha$ Spectropolarimeter \citep[CLASP; see][]{kobayashi11}. The interpretation of the observed spectral line polarization requires the development of suitable modeling tools. To this end, we have developed a three-dimensional (3D), non-LTE multilevel radiative transfer code for modeling the intensity and linear polarization produced by scattering processes in spectral lines and its modification by the Hanle effect. The code is based on accurate and efficient radiative transfer methods (e.g., a 3D formal solver based on short-characteristics with monotonic Bezier interpolation and a highly-convergent iterative method based on the non-linear multigrid method). 

\begin{figure}
\begin{center}
\includegraphics[width=4.8in]{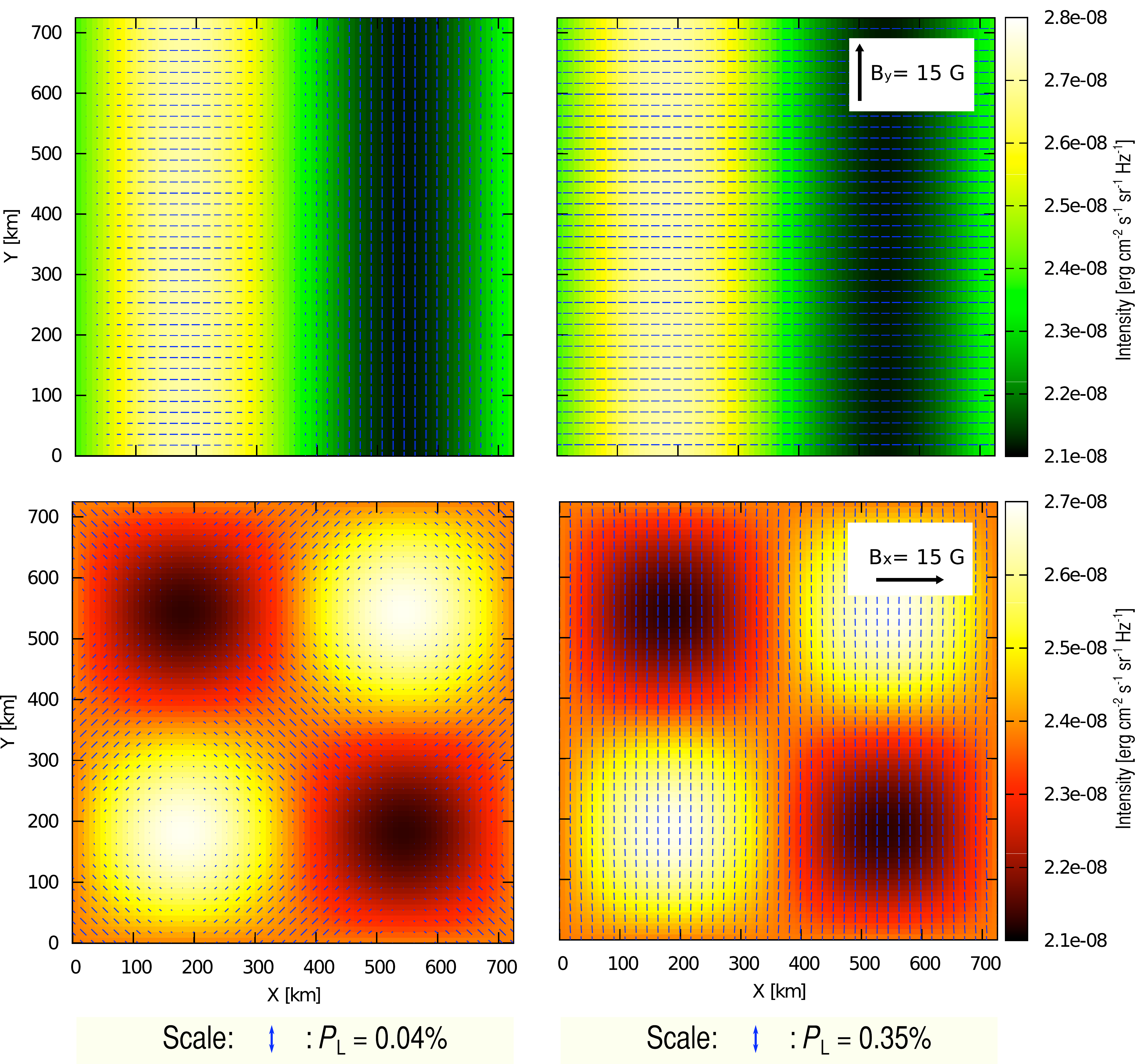}
\end{center}
\caption{Line-center intensity (background color) and linear polarization (arrows) of the hydrogen Ly$\alpha$ line radiation that emerges at the disk center of a solar model atmosphere characterized by sinusoidal temperature fluctuations (amplitude=500 K; horizontal period=725 km) imposed at each height of the 1D semi-empirical model C  of \citet{fal93}. Note the different scales of the polarization vectors in the left (B=0 G) and right (horizontal field with B=15 G) panels.}
\label{stepanjtb:fig:panels}
\end{figure}

Figure~\ref{stepanjtb:fig:panels} shows the results of illustrative 2D (top panels) and 3D (bottom panels) calculations of the linear polarization of the hydrogen Ly$\alpha$ line-center radiation that emerges at the disk center of the chosen solar model. In this forward-scattering geometry (disk center observation) the calculated linear polarization is non-zero in the absence of magnetic fields (left panels), because of the symmetry breaking effects caused by the model's horizontal temperature inhomogeneities \citep[see][]{msjtb11}. The most interesting point to emphasize is that the linear polarization created by the Hanle effect of a horizontal magnetic field of only 15 G is one order of magnitude larger (see right panels). Therefore, its effect can be clearly distinguished.  
Our next step will be to carry out forward modeling calculations in the increasingly realistic 3D atmospheric models described in Carlsson (2012).

\acknowledgements Financial support by the Spanish Ministry of Science and Innovation through project AYA2010-18029 (Solar Magnetism and Astrophysical Spectropolarimetry) and by the grant P209/12/P741 of the Grant Agency of the Czech Republic is gratefully acknowledged.

\end{document}